\documentclass[preprint2]{aastex}

\usepackage{amsmath}
\usepackage{amsfonts}
\usepackage{amssymb}
\usepackage{latexsym}

\newcommand{\be}{\begin{eqnarray}}
\newcommand{\ee}{\end{eqnarray}}
\newcommand{\bse}{\begin{subequations}}
\newcommand{\ese}{\end{subequations}}

\newcommand{\bs}{\boldsymbol}

\newcommand{\mrm}{\mathrm}

\newcommand{\f}{\frac}

\newcommand{\eps}{\epsilon}

\shorttitle{MOND as limit case of DM}
\shortauthors{Dunkel}

\begin{document}

\title{On the relationship between MOND and DM}

\author{J\"orn Dunkel}
\affil{Institut f\"ur Physik, Humboldt-Universit\"at zu Berlin,
  Newtonstra\ss e 15, 12489 Berlin, Germany}
\email{dunkel@physik.hu-berlin.de}

\begin{abstract} 
Numerous astrophysical observations have shown that classical Newtonian
dynamics fails on galactic scales and beyond, if only visible matter is taken
into account. The two most popular theoretical concepts dealing with this
problem are Dark Matter (DM) and Modified Newtonian Dynamics (MOND). In the
first part of this paper it is demonstrated that a generalized MOND equation
can be derived in the framework of Newtonian Dark Matter theory. For systems
satisfying a fixed relationship between the gravitational fields caused by DM
and visible matter, this generalized MOND equation reduces to the traditional
MOND law, first postulated by Milgrom. Therefore, we come to the conclusion
that traditional MOND can also be interpreted as special limit case of DM
theory. In the second part, a formal derivation of the Tully-Fisher relation
is discussed.
\end{abstract}
\keywords{dark matter --- galaxies: kinematics and dynamics}




\section{Introduction}

Seventy years ago, \citet{Zw33,Zw37} was the first to notice
that the speed of galaxies in large clusters is much too great to keep them
gravitationally bound together, unless they are much heavier than one would
estimate on the basis of visible matter. Since those days
numerous further astrophysical observations, e.g., Doppler measurements of
rotation velocities in disk galaxies, have confirmed the failure of the
classical Newtonian theory, if only visible matter is taken into account
\citep{Combes95,Bertin96,Fi99,SaGa02}. Historically, theoretical
concepts addressing this problem can be subdivided in two categories. The first category comprises the Dark Matter (DM) theories \citep{BiTr94,Sa99,Be01,OsSt03}, whereas the second group assumes that Newton's gravitational law requires modification \citep{Mi83a,Mi83b,Mi83c}. 
\par 
DM theories are based on the hypothesis that there exist significant amounts
of invisible (non-baryonic) matter in the universe, interacting with ordinary
visible matter only via gravity. Since empirically very successful, DM has
become a widely accepted cornerstone of the contemporary cosmological standard
model \citep{Sa99,Be01,OsSt03}. Nevertheless, it must also be emphasized that
until now DM has been detected only indirectly by means of its gravitational
effects on the visible matter or the light.  
\par  
Aiming to avoid the introduction of invisible matter, an alternative
phenomenological concept was proposed by \citet{Mi83a,Mi83b,Mi83c}. Instead of adapting the mass distribution, his
approach requires a modified Newtonian dynamics (MOND) in the limit of small
accelerations.  As extensively reviewed by \citet{SaGa02},
this theory can explain galaxy data, such as the flat rotation curves, in a
very 
compelling way. On the other hand, there also have been some indications in
the past that MOND might be an effective or approximate theory, applicable to
a limited range of astrophysical problems only \citep{Ag03}. This hypothesis is supported
by fundamental difficulties associated with relativistic generalizations of
Milgrom's theory \citep{SaGa02,SoWo03,Ag03}. Also, according to \citet{AgScQu01}, MOND seems to become less effective on larger scales; e.g., it cannot account for cluster densities and temperature profiles in detail.
\par
The fact that, to some extend, both DM and MOND can successfully
explain galactic dynamics favors the possibility that there exists a deeper
connection between these two theories [for a general comparison, see
\citep{Ag03}]. Among others, this idea was formulated by \citet{GaBl98b}, and
later pursued by \citet{KaTu02}. Using arguments based on galaxy formation processes in the
early universe, the latter authors claim that MOND follows from cold DM
theory. In his response, \citet{Mi02} questions these results. Among others, he
argues that the predictions made by \citet{KaTu02} would not only conflict with
astronomical observations of pairs of galaxies \citep{GaBl98a}, but also with
numerical results obtained for DM models \citep{NaFrWh97}. Thus, unclarity
still seems  to exist about whether or not MOND can in fact be understood in
the framework of DM \citep{Ag03}.
\par
It is therefore the main purpose of the present paper to explicitely
demonstrate that the MOND equations (if considered as modified Newtonian
gravity) can be derived from classical Newtonian dynamics, provided one also
takes into account the gravitational influence of a DM component. In
particular, it will be shown that the characteristic threshold acceleration,
$a_0\approx 1.2\cdot 10^{-10}$ m/s$^{2}$, below which MOND effects begin to
dominate, can also be interpreted as the asymptotic value of a more general  acceleration field,
characterizing the difference between the gravitational forces caused by
visible matter and dark matter, respectively.

\section{MOND from Newtonian dynamics with DM}

 As starting point, consider the Newtonian EOM of a point-like test particle 
\be\label{e:eom-1}
m\ddot{\bs x}=-m\nabla\left[\Phi_v(\bs x)+\Phi_d(\bs x)\right],
\ee
where $\Phi_v(\bs x)$  and $\Phi_d(\bs x)$ denote the gravitational potentials
due to visible and dark matter, respectively. Both potentials are solutions of
Poisson equations,
\be
\nabla^2\Phi_{v/d}= 4\pi G\;\rho_{v/d},
\ee
where $\rho_{v/d}(\bs x)$ is the corresponding mass density and $G$ denotes the gravitational constant. For convenience, we define the accelerations
\be
\bs g_{v/d}(\bs x):=-\nabla\Phi_{v/d}(\bs x).
\ee
Thus, Eq. \eqref{e:eom-1} simplifies to
\be\label{e:eom-2}
\ddot{\bs x}=\bs g_v + \bs g_d=:\bs g.
\ee
Now let us additionally assume that the acceleration vectors $\bs g_v$ and
$\bs g_v$ point in the same direction, denoted by  
\be\label{e:parallel}
\bs g_v \uparrow\uparrow \bs g_d.
\ee
Note that in this case also $\bs g_{v/d}  \uparrow\uparrow \bs g$. Roughly speaking, the assumptions \eqref{e:parallel} means that the visible mass distribution  $\rho_{v}$ and the DM distribution $\rho_{d}$ behave very similar. Next, we rewrite Eq. \eqref{e:eom-2} as
\be\label{e:eom-3}
\ddot{\bs x}= \left(1 + \f{ g_d}{g_v}\right)\bs g_v,
\ee
where $g_{v/d}:=|\bs g_{v/d}|$ with
\be
g_v
=
g - g_d\ge 0,
\ee
if condition \eqref{e:parallel} holds. Inserting this into \eqref{e:eom-3} yields
\be\label{e:eom-4}
\ddot{\bs x}
&=&
\left(1 + \f{1}{g/g_d-1}\right)\bs g_v .
\ee
Thus, by virtue of \eqref{e:eom-2}, we find that
\be\label{e:neu}
\bs g_v=\left(\f{\eps}{\eps+1}\right)\bs g=: \tilde\mu(\eps)\bs g,
\ee
where we have introduced
\be\label{e:eps}
\eps(\bs x):= \f{g(\bs x)}{g_d(\bs x)}-1\ge 0.
\ee
Equation \eqref{e:neu} can be compared to the fundamental MOND formula \citep{Mi83a,Mi83b,Mi83c,SaGa02} 
\be\label{e:mond}
\bs g_v= \mu\left(\f{g}{a_0}\right)\bs g,
\ee
where, due to empirical reasons, the function $\mu(\xi)$ is {\em postulated} to have the asymptotic behavior
\be\label{e:mond-1}
\mu(\xi)=
\begin{cases}
1,& \xi\gg 1;\\
\xi,&  \xi\ll 1.
\end{cases}
\ee
One readily observes, that this is exactly the {\em natural} asymptotic behavior of $\tilde \mu(\eps)$ for $\eps\to 0$ and $\eps\to \infty$, respectively. Hence, if we identify $\mu$ with $\tilde{\mu}$ and introduce an acceleration field $a(\bs x)$ by
\be\label{e:a(x)}
\f{g(\bs x)}{a(\bs x)}=\eps(\bs x)=\f{g(\bs x)}{g_d(\bs x)}-1,
\ee
then it becomes obvious that \eqref{e:neu} is the natural generalization of the MOND postulate \eqref{e:mond}. The only difference is that we have a local acceleration field $a(\bs x)$ in \eqref{e:neu}, whereas $a_0=const$ was postulated in the MOND formula \eqref{e:mond}. Note, that Eq. \eqref{e:a(x)} can also be written in the equivalent form
\be
\f{1}{a(\bs x)}
&=&\f{1}{g_d(\bs x)}-\f{1}{g(\bs x)}\notag\\
&=&\f{1}{g_d(\bs x)}-\f{1}{g_v(\bs x)+g_d(\bs x)}\label{e:neu-2}.
\ee
Thus, the special MOND case  
\be\label{e:M-limit}
a(\bs x)\equiv a_0
\ee
implies a fixed relation between the acceleration fields due to visible and
dark matter. In particular, since the characteristic MOND acceleration $a_0$
is relatively small, one can further infer from \eqref{e:neu-2} that galaxies
satisfying the MOND limit are DM dominated. 

\section{Axisymmetric disk galaxies and Tully-Fisher law}
\label{s:TF}

In the following, let us concentrate on the quasi-two-dimensional problem of
axisymmetric disk galaxies. It is an experimental observation that for many such systems the Tully-Fisher relation holds \citep{SaGa02,GaBl98a}
\be\label{e:TF-1}
v_\infty^4:=\lim_{r\to\infty} v^4(r)\propto L\propto M,
\ee
where $L$ denotes the luminosity and $M$ is the {\em visible} (baryonic) mass of the galaxy. The quantity $v(r)$ is the absolute velocity of stars or gaseous components,  rotating in the disk plane around the galactic center ($r$ is the distance from the galactic center, defining the origin of the coordinate system). Equating centripetal acceleration $v^2/r$ and $g(r)$, we find
\be\label{e:TF-2}
v^2_\infty=\lim_{r\to\infty} r g(r)=\lim_{r\to\infty} r \sqrt{a(r)\,g_v(r)}.
\ee 
Note that the second equality holds, only if one additionally assumes that
$\eps(r)\ll 1$ for $r\to \infty$. The reason is that, according to
\eqref{e:neu}, only in this very case the approximation $g^2\approx a g_v$ is
valid. Physically, the condition $\eps(r)\ll 1$ reflects a dominating DM
influence, as implied by \eqref{e:eps} and \eqref{e:a(x)}, respectively.
\par
The Tully-Fisher law \eqref{e:TF-1} follows directly from the rhs. of \eqref{e:TF-2}.  
Assuming that $a(r)\to a_\infty$  for $r\to \infty$ and, in agreement with the standard procedure, a Keplerian behavior $g_v(r)\simeq GM/r^2$ for $r\to \infty$, we find the desired result
\be\label{e:TF}
v_\infty^4 =a_\infty G M.
\ee
For the special case $a_\infty=a_0$, this is the well-known MOND formula. Note
that according to our approach Eq. \eqref{e:TF} represents, at least
formally, a derived result, whereas it plays the role of a postulate in
the original MOND papers \citep{Mi83a,SaGa02}.  It might be worthwhile to
emphasize here once again the crucial aspect, which is that the function
$\tilde \mu$ from \eqref{e:neu} naturally satisfies the MOND postulates
\eqref{e:mond-1}. 
\par
Nevertheless, one must be aware of the fact that the above
derivation of Eq. \eqref{e:TF} was essentially guided by the knowledge of the
empirical Tully-Fisher law \eqref{e:TF-1}. More precisely, the DM
paradigm in its current form does {\em  not} provide any explanation for the
fact that in many disk galaxies visible and dark matter have arranged in such a
way that $a(r)$ rapidly converges to a constant non-vanishing value.      
\par
Since $g_v$ and $g_d$ reflect the distributions of visible and dark matter,
and because of
\be\label{e:mond-3}
\f{1}{a_\infty}
=\lim_{r\to \infty}\left\{\f{1}{g_d(r)}-\f{1}{g_v(r)+g_d(r)}\right\},
\ee
the quantity $a_\infty$ gives us information about the asymptotic mass 
distributions.  According to \citep{Mi83a,Mi83b,Mi83c,SaGa02}, for several disk
galaxies the experimental value is given by the MOND value,
$a_\infty=a_0$. From the point of view adopted in this paper, this indicates
that the composition of these galaxies is generally similar. 
\par
In contrast, at least for some clusters of galaxies the actual value of $a(\bs x)$
seems to essentially deviate from the MOND value $a_0$. As mentioned earlier,
 \citet{AgScQu01} have shown that the experimentally
observed, radial temperature profiles of Coma, Abell 2199 and Virgo  can {\em
  not} be fitted if one assumes a globally constant value
$a(\bs x)\equiv a_0$. Furthermore, these authors report satisfactory
agreement when they apply standard DM models instead. With regard to our above
considerations, the latter procedure simply corresponds to using a locally
varying field $a(\bs x)\not\equiv a_0$. On the one hand, this supports the
hypothesis that MOND should be viewed as a special limit case of DM theory; on
the other hand, one is led to ask, why $a(\bs x)$ is approximately constant
in disk galaxies, but seems to vary in clusters. According to the author's
opinion, the answer to this question can only be given by an improved DM
theory, yet to be developed. In particular, such a theory must predict the
dynamics of dark and visible mass components in detail.   
\par
Finally, we still note that if $g_v(\bs x)\ll g_d(\bs x)$ holds, then one can expand \eqref{e:neu-2} yielding
\be
a(\bs x)
\approx 
\f{g_d(\bs x)^2}{g_v(\bs x)}.
\ee
For spherical matter distributions this means that
\be
a(r)\approx \f{[G M_d(r)/ r^{2}]^2}{G M_v(r)/ r^{2}},
\ee
where $M_{v/d}(r)$ denotes the visible/dark mass contained within radius
$r$. For the special case $a(r)\approx a_0$, this is equivalent to 
\be
\f{1}{M_v(r)}\left[\f{M_d(r)}{r}\right]^2
\approx
\f{a_0}{G}
\approx
2\;\f{\mrm{kg\;}}{\mrm{m^2}}
\approx
10^3\;\f{\mrm{M_\odot}}{\mrm{pc^2}},
\ee
which implies a strong correlation between the distributions of visible and
dark matter in the MOND limit. It should be mentioned here that the
possibility of such a connection was already suggested by \citet{GaBl98b}
and, later, also more extensively discussed by \citet{Ga00}.   

\section{Summary and conclusions}

It was shown that the generalized MOND equation \eqref{e:neu} can be derived
from Newtonian dynamics, if one adds a DM contribution  $\Phi_{d}$ to the
(baryonic) Newtonian  potential $\Phi_{v}$, such that $\Phi_{v/d}$ lead to
equally directed accelerations $\bs g_{v/d}=-\nabla \Phi_{v/d}$. Compared to
 the traditional MOND law \eqref{e:mond}, the only formal difference consists in
the fact that the constant threshold value $a_0$ is replaced by the more
general acceleration field $a(\bs x)$ from \eqref{e:neu-2}. In the DM picture, $a(\bs x)$ reflects the local difference between the gravitational forces caused by dark and visible matter, respectively. In order to exactly regain the traditional MOND law \eqref{e:mond}, one additionally has to demand that $a(\bs x)\equiv a_0$. Thus, MOND can in principle also be interpreted as a DM theory, satisfying the two additional conditions \eqref{e:parallel} and \eqref{e:M-limit}.
\par
Therefore, it seems reasonable to assume that the traditional
MOND theory represents a special limit case of Newtonian DM theory. Adopting
this point of view, one can further conclude that MOND successfully explains
the rotation curves of disk galaxies because for such objects the above
conditions \eqref{e:parallel} and \eqref{e:M-limit} are fulfilled. If this is true, then, as also discussed above, the MOND constant $a_0$ can be interpreted as the asymptotic value of the field $a(r)$ as $r\to \infty$. 
\par
More generally speaking, whenever there is a fixed relationship between $g_d$
and $g_v$ (or $\rho_d$ and $\rho_v$, respectively) such that $a(\bs
x)\approx a_0$, then the traditional MOND theory should continue to work
successfully. In turn, if a disk galaxy is in the MOND regime, then
Eq. \eqref{e:neu-2} can be used to estimate the DM distribution $\rho_d$,
provided the visible matter distribution $\rho_v$ is known from
observations. Furthermore, it was shown that $\mu(\xi)=\xi/(\xi+1)$ is the
natural candidate for the MOND function. Another result of this paper was the
formal derivation of the Tully-Fisher law \eqref{e:TF} in
Sec. \ref{s:TF}. This relation should hold whenever the two conditions $g_v\ll
g_d$ and $a_\infty>0$ are satisfied, where $a_\infty:=\lim_{r\to \infty}
a(r)$. In this context it must be stressed that the current DM model cannot explain, in which situations these two conditions are fulfilled, and, if
so, why this is the case. Therefore, modifications of the conventional DM
 theory seem inevitably necessary.      
\par
We conclude this short paper with a more general remark. In principle,
there seems to be an agreement that Newton's theory applied to
visible matter does {\em not} give a generally correct description of the
dynamics of galaxies and, therefore, has to be modified. A first way to do
this is to simply consider an additional potential $\Phi_d$ and, following the
standard strategy, to attach a "generating object" called DM to this
potential. As shown above, Milgrom's concept (if considered as modification of gravity) is in fact very
similar, even though it seems quite different at first glance. In particular,
the MOND equations can also be transformed into a modification of the former
potential type, by starting with $a(\bs x)\equiv a_0$ and reversing the above
manipulations. The "generating object" of the related potential can then be
named DM as well. 

\acknowledgments
The author is very grateful to Christian Theis for his encouraging support and
careful reading of the manuscript. He also wants to thank Stefan Hilbert for
numerous, very helpful discussions and Stacy McGaugh for valuable comments. This
work was, in parts, financially supported by the Studienstiftung des deutschen
Volkes. 


\end{document}